\documentclass[12pt,plain]{article}
\usepackage{listings}
\usepackage[makeroom]{cancel}
\usepackage{fancyhdr}
\usepackage{mathrsfs}
\usepackage{verbatim}
\usepackage{indentfirst}
\usepackage{graphicx}
\usepackage{color}
\usepackage{newlfont}
\usepackage{amssymb}
\usepackage[intlimits]{amsmath}
\usepackage{latexsym}
\usepackage{amsthm}
\usepackage{amscd}
\usepackage{fullpage}
\usepackage{setspace}
\usepackage{multirow}
\usepackage{hyperref}
\usepackage[dvips, hmargin=2.1cm,vmargin=3cm]{geometry}
\usepackage{multirow}
\usepackage{array}
\usepackage{authblk}

\newcommand{\de}{\mathrm{d}}

\newcommand{\Z}{\mathbb{Z}}
\newcommand{\Q}{\mathbb{Q}}
\newcommand{\R}{\mathbb{R}}
\newcommand{\C}{\mathbb{C}}
\newcommand{\A}{\mathcal{A}}

\newcommand{\ha}{\frac{1}{2}}
\newcommand{\tha}{\tfrac{1}{2}}
\newcommand{\qua}{\frac{1}{4}}

\newcommand{\be}{\begin{equation}}
\newcommand{\bey}{\begin{eqnarray}}
\newcommand{\ee}{\end{equation}}
\newcommand{\eey}{\end{eqnarray}}
\newcommand{\ba}{\begin{array}}
\newcommand{\ea}{\end{array}}
\newcommand{\h}{\frak{H}}
\newcommand{\sltr}{\mathrm{SL}(2,\R)}
\newcommand{\tsltr}{\widetilde{\mathrm{SL}}(2,\R)}

\newcommand{\sltz}{\mathrm{SL}(2,\Z)}
\newcommand{\Hei}{\mathbb{H}(\mathbb{R})}
\newcommand{\ve}[2]{\left(\ba{c}#1\\#2\ea\right)}

\newcommand{\ma}[4]{\left(\ba{cc}#1&#2\\#3&#4\ea\right)}

\newcommand{\bm}[1]{\mbox{\boldmath{$#1$}}}

\newcommand{\sve}[2]{\left(\begin{smallmatrix}\!#1\!\\ \!#2\!\end{smallmatrix}\right)}
\newcommand{\sma}[4]{\left(\begin{smallmatrix} #1&#2\\#3&#4\end{smallmatrix}\right)}
\newcommand{\e}[1]{e\!\left(#1\right)}

\def\GamG{\Gamma\backslash G}

\def\gene{\gamma}
\def\ASL{\operatorname{ASL}}

\def\vecxi{{\text{\boldmath$\xi$}}}

\def\vepsi{{\text{\boldmath$\psi$}}}
\def\vePsi{{\text{\boldmath$\Psi$}}}

\def\vecnull{{\text{\boldmath$0$}}}

\theoremstyle{plain}
\newtheorem{theorem}{Theorem}[section]
\newtheorem{lem}[theorem]{Lemma}

\theoremstyle{definition}

\newtheorem{remark}[theorem]{Remark}

\input xy
\xyoption{all}

\begin{document}
\title{Autocorrelation functions for quantum particles \\in supersymmetric P\"{o}schl-Teller Potentials}
\author{F. Cellarosi$^*$}
\date{ }
\maketitle

\let\oldthefootnote\thefootnote
\renewcommand{\thefootnote}{\fnsymbol{footnote}}
\footnotetext[1]{Queen's University. Department of Mathematics and Statistics. Jeffery Hall, 48 University Avenue,
Kingston, ON Canada, K7L 3N6. \texttt{fc19@queensu.ca}}
\let\thefootnote\oldthefootnote

\begin{abstract}
We consider autocorrelation functions for supersymmetric quantum mechanical systems (consisting of a fermion and a boson) confined in trigonometric P\"oschl-Teller partner potentials. We study the limit of rescaled autocorrelation functions (at random time) as the localization of the initial state goes to infinity. The limiting distribution  can be described using pairs of Jacobi theta functions on a suitably defined homogeneous space, as a corollary of the work of Cellarosi and Marklof. 
A construction by Contreras-Astorga and Fern\'andez, 
provides large classes of P\"oschl-Teller partner potentials  to which our analysis applies
\end{abstract}


\section{Introduction}
\subsection{Supersymmetric quantum mechanics}
\emph{Supersymmetric (SUSY) quantum mechanics } is the study of a pair of Hamiltonians (in units where $\hbar=\mbox{mass}=1$)
\begin{align}
H_0=-\frac{1}{2}\frac{\de^2}{\de x^2}+V_0(x),&&H_1=-\frac{1}{2}\frac{\de^2}{\de x^2}+V_1(x)\label{eq:two-Hamiltonians}
\end{align}
 that are intertwined by a differential operator $A$ and its adjoint $A^\dag$ as
\begin{align}
H_0A=AH_1,&&H_1A^\dag=A^\dag H_0. \label{eq:intertwining}
\end{align}
The term `supersymmetric' is motivated as follows. If we define the operator matrices
\begin{align}
&\mathbf{Q}=\left(\begin{array}{cc}0 & 0 \\A & 0\end{array}\right),&&\mathbf{Q}^\dag=\left(\begin{array}{cc}0 & A^\dag \\0 & 0\end{array}
\right),&&\mathbf{H}_{\mathrm{ss}}=\left(\begin{array}{cc}A^\dag A & 0 \\0 & A A^\dag\end{array}\right),\\
&\mathbf{Q}_1=\frac{\mathbf{Q}^\dag+\mathbf{Q}}{\sqrt2},&&\mathbf{Q}_2=\frac{\mathbf{Q}^\dag-\mathbf{Q}}{i\sqrt{2}},&&
\end{align}
then we have the commutator and anticommutator relations of the \emph{supersimmetry algebra with $N=2$ generators} 
\begin{align}
[\mathbf{Q}_1,\mathbf{H}_{\mathrm{ss}}]=0,&&[\mathbf{Q}_2,\mathbf{H}_{\mathrm{ss}}]=0,&& \{\mathbf{Q}_1,\mathbf{Q}_1\}=\{\mathbf{Q}_2,\mathbf{Q}_2\}=\mathbf{H}_{\mathrm{ss}},&&\{\mathbf{Q}_1,\mathbf{Q}_2\}=0,\label{SUSY-algebra}
\end{align}
where $[X,Y]=XY-YX$ and $\{X,Y\}=XY+YX$ denote the commutator and the anticommutator of $X$ and $Y$, respectively. 
The algebra \eqref{SUSY-algebra} corresponds to the simplest supersymmetric quantum system (see \cite{MR683171}, as well as \S5 of \cite{MR2100344}).
The operators $\mathbf{Q}, \mathbf{Q}^\dag$ are called \emph{supercharges} or \emph{supersymmetry generators}, and the Hamiltonians $H_0, H_1$ (as well as  the potentials $V_0, V_1$) are called \emph{supersymmetric partners}.

Given a general 1-dimensional Hamiltonian $ H_0$ whose eigenfunctions and eigenvalues are known, there is an intertwining   method due to Sukumar (see  \cite{SUSYQM-Fernandez} for a historical account, tracing back to the work of Dirac)  to construct various partners $H_1$ using various differential operators $A,A^\dag$. The advantage of  the intertwining relations \eqref{eq:intertwining} is to generate eigenvalues and eigenfunctions of $H_1$ from those of $H_0$. 
In general, if $A$ is a differential operator of order $k$, then the spectra of $H_0$ and $H_1$ differ by at most $k$ values. Moreover, if we denote by
\begin{align}
\mathbf{H}=\left(\begin{array}{cc}H_1 & 0 \\0 & H_0\end{array}\right).\label{eq:physical-hamiltonian}
\end{align}
 the \emph{physical Hamiltonian}, then the \emph{supersymmetric Hamiltonian} $\mathbf{H}_{\mathrm{ss}}$  in \eqref{SUSY-algebra} can be expressed as a polynomial of degree $k$ in $\mathbf{H}$. It is therefore enough to study the time-independent Schr\"odinger equation 
\begin{align}
\mathbf{H}\vepsi=E\vepsi,\label{eq:ti-schrodinger-eq}
\end{align}
where $\vepsi(x)=\left(\begin{array}{c}\psi_1(x) \\\psi_0(x)\end{array}\right)$. 
From the physical point of view, SUSY predicts that each particle of the Standard Model has a partner particle with a spin that differs by half unit. Therefore, we may think of $\mathbf H$ as describing the joint action of $H_0$ on fermions (particles with half-integer spin) and of $H_1$ on bosons (particles with integer spin).
\subsection{Autocorrelation functions}
The time-dependent Schr\"odinger equation corresponding to \eqref{eq:ti-schrodinger-eq} is 
\begin{align}
i\frac{\partial}{\partial t}\vePsi=\mathbf{H}\vePsi,\label{eq:td-schrodinger-eq}
\end{align}
where $\vePsi(x,t)=\left(\begin{array}{c}\Psi_1(x,t) \\\Psi_0(x,t)\end{array}\right)$.
We  consider the 1-dimensional case in which $x\in I$, where $I\subseteq \R$ denotes an interval. Moreover, we  restrict our analysis to the class of physical Hamiltonians $\mathbf{H}$ acting on $\mathrm{L}^2(I)\oplus\mathrm{L}^2(I)$ with purely discrete spectrum. This means that there exists a orthonormal basis  of $\mathrm{L}^2(I)\oplus\mathrm{L}^2(I)$ consisting of eigenfunctions of $\mathbf{H}$, say $\vepsi_n(x)=\left(\begin{array}{c}\psi_{1,n}(x) \\\psi_{0,n}(x)\end{array}\right)$ with eigenvalues $E_n$, $n\geq0$. In this case the solution of \eqref{eq:td-schrodinger-eq} with initial condition $\vePsi(x,0)$ is given in terms of the evolution operator $\mathbf{U}(t)=e^{-i \mathbf{H}t}$ as 
\begin{align}
\vePsi(x,t)=\mathbf{U}(t)\vePsi(x,0)=\sum_{n=0}^\infty \mathbf{c}_n e^{-i E_n t}\vepsi_n(x),
\end{align}
where 
\begin{align}
\textbf{c}_n=\left(\begin{array}{c}c_{1,n}\\c_{0,n}\end{array}\right),\hspace{.5cm}c_{\ell,n}=\int_I \Psi_\ell(x,0), \overline{\psi_{\ell,n}(x)}\,\de x,\hspace{.5cm}\ell=0,1.
\end{align}
The \emph{autocorrelation function} for the initial state $\vePsi(x,0)$ is the function $\textbf A:\R\to\C^2$,
\begin{align}
\textbf A(t)=\int_{I}\vePsi(x,0)\overline{\vePsi(x,t)}\,\de x=\left(\begin{array}{c}\displaystyle\sum_{n\geq 0} |c_{1,n}|^2 e^{i E_n t} \\\displaystyle \sum_{n\geq 0} |c_{0,n}|^2 e^{i E_n t}\end{array}\right).\label{def-A(t)}
\end{align}
We further restrict our attention to  the case of 
autocorrelation functions $\mathbf A(t)$ as in \eqref{def-A(t)} where the coefficients $c_{\ell,n}$ have a special form, namely 
\begin{align}
\left|c_{\ell,n}\right|^2=\frac{1}{N}f_\ell\!\left(\frac{n}{N}\right),\end{align} for some nonnegative functions $f_1,f_0$ on $\R_{\geq0}$ and some $N\geq 1$.
This is a natural choice, which appears in several interesting cases. 
Define %
\begin{align}
\mathbf A_N(f_1,f_0;t)=\left(\begin{array}{c}\displaystyle\frac{1}{N}\sum_{n\geq 0}f_1\!\left(\frac{n}{N}\right)e^{i E_n t} \\ \displaystyle\frac{1}{N}\sum_{n\geq 0}f_0\!\left(\frac{n}{N}\right)e^{i E_n t}\end{array}\right)
\end{align}
We will consider the \emph{large $N$} regime, which physically represents the case of initial states $\vePsi(x,0)$ that are highly localized in space. We will focus on the autocorrelation functions for a class of Hamiltonians $\textbf{H}$ as in \eqref{eq:physical-hamiltonian} where the spectra of $H_0$ and of $H_1$ differ by at most finitely many values.  
For random $t$, we will treat $\mathbf A_N(f_1,f_0;t)$ as a random variable at the scale $N^{-1/2}$. In this case, as $N\to\infty$ the contribution of finitely many eigenvalues to $\mathbf A_N(f_1,f_0, t)$ is $O(N^{-1})$ and hence negligible in our analysis. Therefore, without loss of generality, we can assume that $H_0$ and $H_1$  have the same spectrum, and thus $\mathrm{sp}(\mathbf H)=\mathrm{sp}(H_0)=\mathrm{sp}(H_1)$. 
%
%
%
\subsection{Trigonometric P\"{o}schl-Teller potentials}
Let us discuss the case when $H_0$ is the Hamiltonian corresponding to the so-called \emph{trigonometric P\"{o}schl-Teller} potential 
with parameters $(\alpha,\beta)$, i.e.
\begin{align}
V_0(x)=\frac{(\alpha-1)\alpha}{2\sin^2(x)}+\frac{(\beta-1)\beta}{2\cos^2(x)},\hspace{.5cm}\alpha,\beta>1,\hspace{.5cm} 0\leq x\leq \frac{\pi}{2}.\label{def-V0}
\end{align}
The potential $V_0$ can be interpreted as an infinite well, confining the particle to the 1-dimensional box $[0,\frac{\pi}{2}]$, with ``soft walls'', where the the parameters $\alpha$ and $\beta$ represent the strength of the reflection of the particle off the two walls.
Denote $\gamma=\alpha+\beta$.
The eigenvalues of $H_0$ on $\mathrm L^2([0,\frac{\pi}{2}])$ are 
\begin{align}\label{eigenvalues-of-H0}
E_n=\frac{1}{2}(2n+\gamma)^2,\hspace{.5cm}n\geq0.
\end{align}
The corresponding normalized eigenfunctions are
\begin{align}
\psi_{0,n}(x)=
\sqrt{\frac{2(2n+\gamma)\Gamma(n+\gamma)(\alpha+\frac{1}{2})_n}{n!\,\Gamma(\alpha+\frac{1}{2})\Gamma(\beta+\frac{1}{2})(\beta+\frac{1}{2})_n}}
\sin^{\alpha}(x)\cos^{\beta}(x)\,\,_2F_1\!\left(-n,n+\gamma;\alpha+\tfrac{1}{2};\sin^2(x)\right),
\end{align}
where $_2F_1(a,b;c;z)=\sum_{n=0}^\infty\frac{(a)_n (b)_n}{(c)_n}\frac{z^n}{n!}$ is the hypergeometric function, and $(a)_n=\frac{\Gamma(a+n)}{\Gamma(a)}$ is 
the Pochhammer symbol. See \cite{MR2515866} and \cite{dutt-khare-sukhatme}.
 
\begin{figure}[h!]
\begin{center}
\includegraphics[scale=.63]{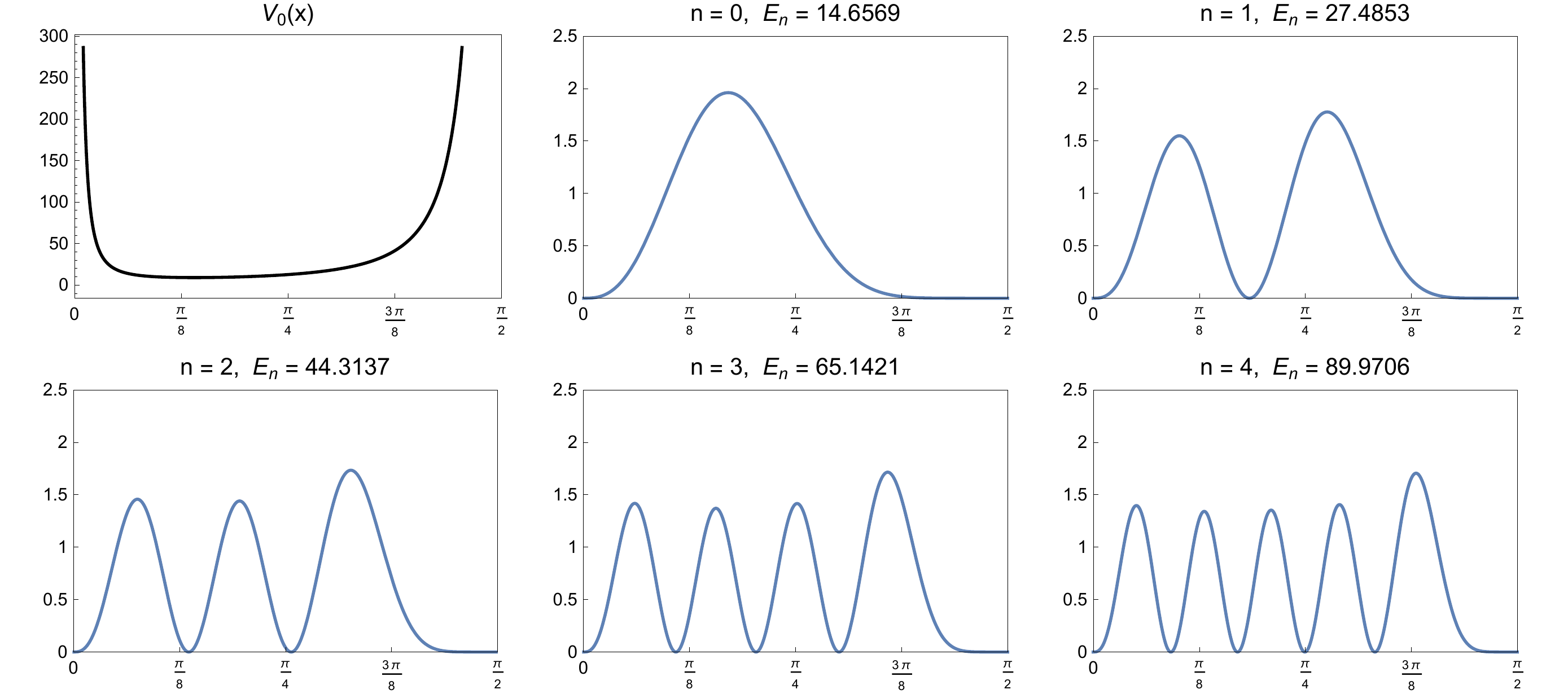}
\caption{The potential $V_0(x)$ for $\alpha=\sqrt2$ and $\beta=4$ and the probability densities $x\mapsto |\psi_{0,n}(x)|^2$ on $[0,\frac{\pi}{2}]$ for $0\leq n\leq 4$.}
\label{six-eigenfunctions}
\end{center}
\end{figure}

Using various differential operators  $A,A^\dag$ as in \eqref{eq:intertwining}, 
Contreras-Astorga and Fern\'andez \cite{MR2515866}, were able to construct various families of partner potentials $V_1$ such that $H_0$ and $H_1$ are isospectral. We  discuss two such families in Section  \ref{section-many-susy-partners}.

\subsection{The main theorem}\label{subsection-main-theorem}
Let $H_0=-\frac{1}{2}\frac{\de^2}{\de x^2}+V_0$ be the Hamiltonian with with P\"oschl-Teller potential \eqref{def-V0} with parameters $(\alpha,\beta)$. Let $\gamma=\alpha+\beta$. Let $H_1$ be any supersymmetric partner of $H_0$ such that $\mathrm{sp}(H_1)$ and $\mathrm{sp}(H_0)$ differ by at most finitely many eigenvalues. Fix two compactly supported, Riemann integrable  functions $f_0,f_1:\R_{\geq0}\to\R$. 
Let $\lambda$ be a probability measure on $\R$, absolutely continuous with respect to the Lebesgue measure. Denote by $\rho$ the density of $\lambda$, i.e. $\de\lambda(t)=\rho(t)\de t$. Rescaled autocorrelation functions are viewed as random variables, i.e. as $\C^2$-valued measurable functions of $t\in(\R, \mathscr{B}(\R), \lambda)$. More precisely, consider the random variables 
\begin{align}\label{def_X_N}
\mathbf X_{N}: (\R,\mathscr{B}(\R),\lambda)\longrightarrow (\C^2,\mathscr{B}(\C^2)),\hspace{1cm}\mathbf X_N(t)=\sqrt{N}\mathbf A_N(f_0,f_1;t),\hspace{1cm} N\geq1.
\end{align}
 Throughout the paper, $\mathscr{B}(\cdot)$ denotes the Borel $\sigma$-algebra. Our main result is the following
\begin{theorem}\label{main-thm}
Assume $\gamma\notin\mathbb Q$. Then the random variables $\mathbf X_N$ have a limiting distribution as $N\to\infty$. That is, 
there exists a (non-degenerate) random variable $\mathbf X$ on $\C^2$ such that $\mathbf X_N$ converge in law to $\mathbf X$ as $N\to\infty$. Moreover, the law of $\mathbf X$ (which does not depend on $\rho$ nor $\gamma$) can be realized as the push forward onto $\C^2$ of the Haar measure on a homogeneous space $\Gamma\backslash G$ via an explicit function $\Gamma\backslash G\to\C^2$. 
\end{theorem}

\begin{figure}[h!]
\hspace{-2cm}
\includegraphics[scale=.9]{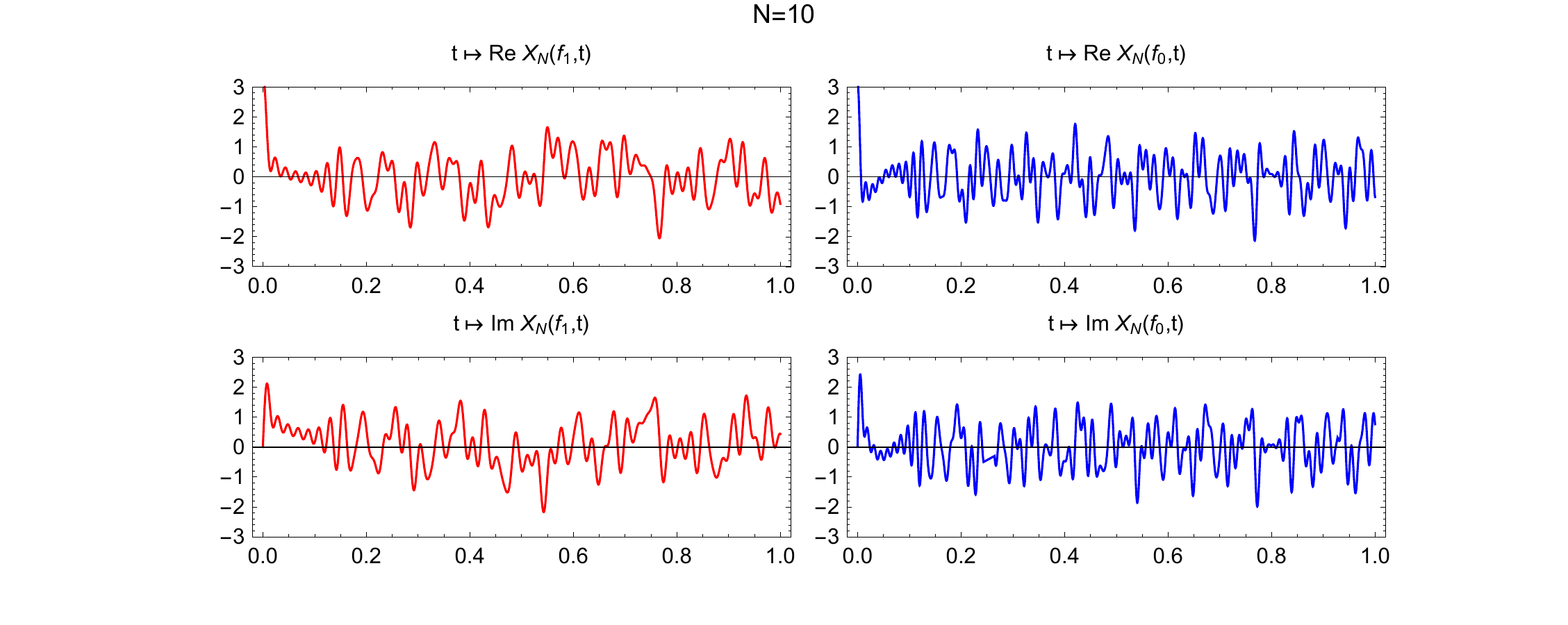}

\hspace{-2cm}
\includegraphics[scale=.9]{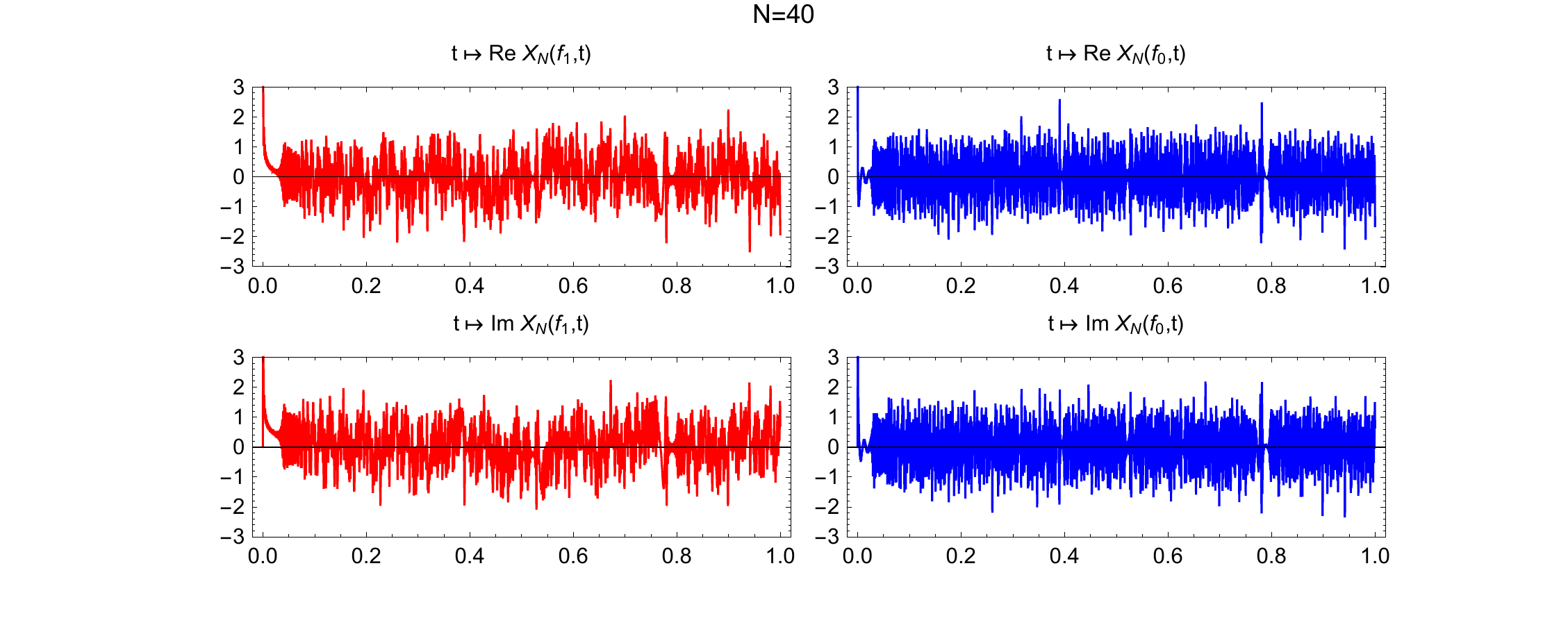}
%
\caption{The 
 real and imaginary parts of the two components of the rescaled autocorrelation function $t\mapsto\mathbf{X}_N(t)=(X_N(f_1,t),X_N(f_0,t))$ 
 for $f_1=\mathbf{1}_{[0,1]}$ (left panels) and $f_0=\mathbf{1}_{\left[\tfrac{1}{3},\tfrac{4}{3}\right]}$ (right panels) and $N=10$ (top four panels) and $N=40$ (bottom four panels). In all the panels, $\alpha=\sqrt{2}$ and $\beta=3$. 
}\label{figure-N10-N50} 
\end{figure}

\begin{remark}
 In this case $\Gamma<G$ is a lattice in the Lie group $G$, and $\Gamma\backslash G$ has finite volume. We  describe $\Gamma\backslash G$, its normalized Haar measure, and the function $\mathbf{X}$ on $\Gamma\backslash G$ explicitly in Section \ref{section-X}. The proof of Theorem \ref{main-thm} is provided in Section \ref{section-proofs}.

In particular we show that the law of $\mathbf X$ is  \emph{not} the product of two measures on $\C$. This means that the two random variables  \begin{align}\label{two-RVs}
\displaystyle\frac{1}{\sqrt N}\sum_{n\geq 0}f_0\!\left(\frac{n}{N}\right)e^{i E_n t}\hspace{.3cm}\mbox{and}\hspace{.3cm}\displaystyle\frac{1}{\sqrt N}\sum_{n\geq 0}f_1\!\left(\frac{n}{N}\right)e^{i E_n t}
\end{align} do not become independent as $N\to\infty$. This can be interpreted physically as follows. The probability distribution of the autocorrelation function (at random time) of an individual particle is altered if we condition on the fact that the autocorrelation function of the partner particle lies in a certain set in $\C$, at least for highly localized particles. This means, in particular, that if we  observe --at a random time $t$-- a localized quantum particle undergoing a quantum revival (i.e. we observe a large value of its correlation function), then the probability that the partner particle (which we may have no access to) is also undergoing a quantum revival at time $t$ is not the same as the unconditional probability of the same event, see Remark \ref{remark-independence} 
\end{remark}
As an illustration, In Figure \ref{figure-N10-N50} we plot the real and imaginary parts of both components of $\mathbf{X}_N(t)$ when $f_1$ and $f_0$ are indicator functions, for different values of $N$.

\section{The limiting random variable $\mathbf X$}\label{section-X}
In this section we describe explicitly the random variable $\mathbf X$ featured in Theorem \ref{main-thm}.
 We refer to  \S2 of \cite{PLMS:PLMS0775} for more details. 

\subsection{The universal Jacobi group $G$}\label{section-universal-Jacobi}
Let $\h:=\{w\in\C:\:\Im(w)>0\}$ denote the upper half plane. 
The group $\sltr$ acts on $\h$ by M\"{o}bius transformations $z\mapsto gz:=\frac{az+b}{cz+d}$, where $g=\sma{a}{b}{c}{d}\in\sltr$. 
Every $g\in\sltr$ can be written uniquely via Iwasawa decomposition as
\be g=n_xa_yk_\phi ,\label{Iwasawa-dec}\ee
where 
\be n_x=\ma{1}{x}{0}{1},\:\:a_y=\ma{y^{1/2}}{0}{0}{y^{-1/2}},\:\:k_\phi=\ma{\cos\phi}{-\sin\phi}{\sin\phi}{\cos\phi},
\label{def-of-iwasawa-pieces}\ee
and $z=x+i y\in\h$, $\phi\in[0,2\pi)$. 
%
%
%
%
Set $\e{t}:=e^{2\pi i t}$ and $\epsilon_g(z)=(cz+ d)/|cz+ d|$. The universal cover of $\sltr$ is defined as 
\be\tsltr:=\{[g,\beta_g]:\:g\in\sltr,\:\beta_g \mbox{ a continuous function on $\h$ s.t. }\e{i\beta_g(z)}=\epsilon_g(z)\},\ee 
and has the group structure given by
\be[g,\beta_g^1][h,\beta_h^2]=[gh,\beta_{gh}^3],\qquad\beta_{gh}^3(z)=\beta_g^1(hz)+\beta_h^2(z),\label{mult-tsltr}
\ee
\be
[g,\beta_g]^{-1}=[g^{-1},\beta_{g^{-1}}'],\qquad \beta_{g^{-1}}'(z)=-\beta_{g}(g^{-1}z).\ee
$\tsltr$ is identified with $\h\times\R$ via $[g,\beta_g]\mapsto(z,\phi)=(g i,\beta_g(i))$ and it acts on $\h\times\R$ via
\be [g,\beta_g](z,\phi)=(gz,\phi+\beta_g(z)).\label{act-tsltr-on-hxR}\ee 

The Iwasawa decomposition \eqref{Iwasawa-dec} of $\sltr$ extends to a decomposition of $\tsltr$: 
for every $\tilde g=[g,\beta_g]\in\tsltr$ we have
\be\tilde g=[g,\beta_g]=\tilde n_x\tilde a_y\tilde k_\phi=[n_x,0][a_y,0][k_\phi,\beta_{k_\phi}].\label{Iwasawa-tlstr}\ee
%
%
Let $\omega$ be the standard symplectic form on $\R^2$, $\omega(\vecxi,\bm{\xi'})=xy'-yx'$, where $\vecxi=\sve{x}{y},\bm{\xi'}=\sve{x'}{y'}$. The Heisenberg group $\Hei$ is defined as $\R^2\times\R$ with the multiplication law
\be(\vecxi,t)(\bm{\xi'},t')=\left(\vecxi+\bm{\xi'},t+t'+\tfrac{1}{2}\omega(\vecxi,\bm{\xi'})\right).\ee
%
%
We consider universal Jacobi group 
 \be G=\tsltr\ltimes\Hei 
 ,\ee
having the multiplication law
\be([g,\beta_g];\vecxi,\zeta)([g',\beta'_{g'}];\bm{\xi'},\zeta')=\left([gg',\beta''_{gg'}];\vecxi+g\bm{\xi'},\zeta+\zeta'+\tha\omega(\vecxi,g\bm{\xi'})\right)\label{mult--univ-Jacobi},\ee
where $\beta''_{gg'}(z)=\beta_g(g'z)+\beta'_{g'}(z)$.
The Haar measure on $G$ is given in coordinates $(x+i y,\phi;\scriptsize{\sve{\xi_1}{\xi_2}},\zeta)$ by \be\de \mu(g)=\frac{\de x\,\de y\,\de\phi\,\de\xi_1\,\de\xi_2\,\de\zeta}{y^2}.\label{Haar-measure-on-G}\ee

\subsection{Hermite expansion}
Let  $H_k$ be the $k$-th Hermite polynomial
\begin{align} 
H_k(t)&=(-1)^k e^{t^2}\frac{\de^k}{\de t^k}e^{-t^2}
= k! \sum_{m=0}^{\lfloor{\frac{k}{2}}\rfloor} \frac{(-1)^m(2t)^{k-2m}}{m!(k-2m)!}.
\end{align}
%
We define the  Hermite functions 
\begin{align}
\psi_k(t)=\left(2\pi\right)^{\qua}h_k(\sqrt{2\pi}t)=(2^{k-\ha}k!)^{-1/2} H_k(\sqrt{2\pi}\,t)e^{-\pi t^2},
\end{align}
which form an orthonormal basis for $L^2(\R,\de x)$. Let $f\in \mathrm{L}^2(\R)$. 
Following \cite{Marklof-1999},  we  set 
\be f_\phi(t)=\sum_{k=0}^\infty\hat f(k)e^{- i (2k+1)\phi/2}\psi_k(t),
\label{f_phi-Hermite}\ee
where
$\hat f(k)=\langle f, \psi_k \rangle_{L^2(\R)}$. For an equivalent definition of $f_\phi$, see \S2.3 of \cite{PLMS:PLMS0775}. 
The space of functions $f:\R\to\R$ for which $f_\phi$ has a prescribed decay at infinity, uniformly in $\phi$, is denoted by $\mathcal S_\eta(\R)$. More precisely
\be\mathcal S_{\eta}(\mathbb{R}):=\left\{f:\R\to\R:\:\:\:\sup_{t,\phi}|f_\phi(t)|(1+|t|)^\eta<\infty\right\},\label{def-kappa-eta}\ee
see e.g. \cite{Marklof2007b}. 

\subsection{Jacobi theta functions on $G$}
For $g=(z,\phi;\vecxi,\zeta)\in G$ and $f\in\mathcal S_\eta(\R)$, $\eta>1$ 
define the \emph{Jacobi theta function} as
\be
\Theta_f(z,\phi;\vecxi,\zeta)=y^{1/4}e(\zeta-\tha \xi_1\xi_2)\sum_{n\in\Z}f_{\phi}\left((n-\xi_2)y^{1/2}\right)e\!\left(\tha(n-\xi_2)^2x+n\xi_1\right)\label{Jacobi-theta-sum-2},\ee
where $z=x+ i y$, $\vecxi=\sve{\xi_1}{\xi_2}$ and $f_\phi$ is as in \eqref{f_phi-Hermite}. 

\subsection{A lattice $\Gamma<G$ such that $\Theta_f$ is $\Gamma$-invariant}
Consider the following elements of $G$, each written in two ways using the identification described in Section \ref{section-universal-Jacobi}):
\begin{align}
\gene_1&=\left(\left[\ma{0}{-1}{1}{0},\arg\right];\bm 0,\frac{1}{8}\right)=\left(i,\frac{\pi}{2};\bm0,\frac{1}{8}\right),\label{invariance-by-h1}\\
\gene_2&=\left(\left[\ma{1}{1}{0}{1},0\right];\ve{1/2}{0},0\right)=\left(1+i,0;\ve{1/2}{0},0\right),\\
\gene_3&=\left(\left[\ma{1}{0}{0}{1},0\right];\ve{1}{0},0\right)=\left(i,0;\ve{1}{0},0\right),\\
\end{align}
\begin{align}
\gene_4&=\left(\left[\ma{1}{0}{0}{1},0\right];\ve{0}{1},0\right)=\left(i,0;\ve{0}{1},0\right),\\
\gene_5&=\left(\left[\ma{1}{0}{0}{1},0\right];\ve{0}{0},1\right)=\left(i,0;\bm 0,1\right).
\end{align}
It was shown in \cite{Marklof2003ann} that for $i=1,\ldots,5$ we have
$\Theta_f(\gene_i g)=\Theta_f(g)$ for every $g\in G$.
The Jacobi theta function $\Theta_f$ is therefore invariant under the left action by the group 
\be \Gamma=\langle \gene_1,\gene_2,\gene_3,\gene_4,\gene_5\rangle
<G,\label{def-Gamma}.\ee 
This means that $\Theta_f$ is well defined on the quotient $\Gamma\backslash G$. 
%
The group 
$\Gamma$ is a lattice in $G$ and 
$\Gamma\backslash G$ is a 4-torus bundle over the modular surface $\sltz\backslash\frak H$. In particular $\GamG$ is non-compact. A fundamental domain for the action of $\Gamma$ on $G$ is 
\be \mathcal F_{\Gamma}=\left\{(z,\phi;\vecxi,\zeta)\in\mathcal F_{\sltz} \times [0,\pi) \times [-\tfrac12,\tfrac12)^2\times [-\tfrac12,\tfrac12) 
\right\},\label{fund-dom-H}\ee
where $\mathcal F_{\sltz}$ is a fundamental domain of the modular group in $\h$. The hyperbolic area of $\mathcal F_{\sltz}$ is $\frac{\pi}{3}$, and hence,
by \eqref{Haar-measure-on-G}, we have that $\mu(\GamG)=\mu(\mathcal F_\Gamma)=\frac{\pi^2}{3}$.

\begin{remark}
Although we defined  the Jacobi theta function in \eqref{Jacobi-theta-sum-2} assuming that $f$ is regular enough ($\eta>1$), it can be shown 
$\Theta_f$ is a well defined element of $L^2(\GamG,\mu)$  (in fact, of $L^4(\GamG,\mu)$) provided $f\in L^2(\R)$, see \S2.9 of \cite{PLMS:PLMS0775}.
\end{remark}

We are finally ready to define $\mathbf X$. Given $f_0,f_1$ as in the Section \ref{subsection-main-theorem}, set
$\mathbf{X}: \Gamma\backslash G\to\C^2$ as 
\begin{align}
\mathbf{X}(\Gamma g)=\ve{\Theta_{f_1}(\Gamma g)}{\Theta_{f_0}(\Gamma g)}.
\end{align}
Observe that $\mathbf X:\left(\GamG,\mathscr{B}(\GamG),\frac{3}{\pi^2}\mu\right)\to(\C^2,\mathscr{B}(\C^2))$
 is a random variable whose law is simply the push forward of the normalized Haar measure $\frac{3}{\pi^2}\mu$ onto $\C^2$ via $\mathbf{X}$.
The properties of the law of each component $\Theta_{f}(\Gamma g)$ when $\Gamma g$ is Haar-random on $\GamG$ have been studied in \cite{PLMS:PLMS0775}. In particular, we have the following
\begin{lem}[\cite{PLMS:PLMS0775}]\label{lemma-tail}
Let $\eta>1$ and $f\in\mathcal{S}_\eta$. 
Then for all sufficiently large $R>0$ we have 
\be\label{mu(tail>R)-formula}
\mu(\{g\in\GamG:\:|\Theta_f(g)|>R\})= \frac{2}{3}D(f) R^{-6} \big(1+O_\eta(\kappa_\eta(f)^{2\eta} R^{-{2\eta}}) \big),
\ee
where $D(f):=\displaystyle\int_{-\infty}^\infty\int_0^{\pi}|f_{\phi}(w)|^6\de\phi\,\de w$ and $\kappa_\eta(f)=\displaystyle\sup_{w,\phi}|f_\phi(w)|(1+|w|)^\eta.$ 
\end{lem} 
If $f$ is less regular (e.g. when $f$ is the indicator of an interval, in which case $f\in\mathcal S_1$ and Lemma \ref{lemma-tail} does not apply) a delicate analysis can be performed to study the tail asymptotic (see \S3.6 of \cite{PLMS:PLMS0775}). In any case, the complex-valued random variable $\Theta_f$ has heavy tails and  all moments of order $p\geq 6$  are infinite.

\section{Proof of Theorem \ref{main-thm}} 
 \label{section-proofs}
Recall \eqref{Iwasawa-tlstr}. The following theorem
describes the how  certain ``irrational'' horocycle  lifts of the form  $u\mapsto M(u)\Psi^u$, with $M(u)\in G$ and $\Psi^u=(\tilde n_u,\sve{0}{0},0)$, become equidistributed in $\GamG$ under the action of the geodesic flow $\Phi^s=(\tilde a_{e^{-s}},\sve{0}{0},0)$. It follows immediately from Corollary  4.3 in \cite{PLMS:PLMS0775}. 
 \begin{theorem}\label{thmJEQ2}
Let $\sigma:\R\to\R_{\geq0}$ be a probability density, and let $F:\GamG\to\C^2$ be a bounded continuous function. For any 
$\theta\in \R\smallsetminus\Q$ and any $\varpi\in\R$ 
we have 
\begin{equation}\label{statement-limit-thm}
\lim_{s\to\infty} \int_{\R} F\!\left(\Gamma\!\left(I_2; \sve{\theta u}{0},  \varpi u \right)\Psi^u \Phi^s\right) 
\sigma(u)\de u = \frac{1}{\mu(\GamG)} \int_{\GamG} F(\Gamma g) \de\mu(\Gamma g) .
\end{equation}
\end{theorem}
Note that the limit does not depend on neither $\sigma$, nor  $\theta$ (provided it is irrational), nor $\varpi$.
Our assumptions on the P\"{o}schl-Teller potential $V_0$ and on the functions $f_0,f_1$, along with \eqref{eigenvalues-of-H0}, allow us to write $\tfrac{1}{2\pi}E_n t=\tfrac{1}{2\pi}\tfrac{(2n+\gamma)^2}{2}t=\frac{1}{2}(n-\xi_2)^2u+n\xi_1+\zeta-\frac{1}{2}\xi_1\xi_2$, where $\xi_1=\frac{\gamma}{2}u$, $\xi_2=0$,  $\zeta=\frac{\gamma^2}{8}u$, and  $u=\frac{2t}{\pi}$. Therefore, using \eqref{Jacobi-theta-sum-2} and setting $s=2\log N$, we have
\begin{align}
\frac{1}{\sqrt{N}}\sum_{n\geq0} {f_\ell}\!\left(\frac{n}{N}\right) e^{i E_n t}=
\Theta_{f_\ell}\!\left(u+ie^{-s},0;\sve{\frac{\gamma}{2}u}{0},\tfrac{\gamma^2}{8}u\right), \hspace{.3cm}\ell=0,1.
\end{align} 

Recall that $\rho$ denotes the density of the probability measure $\lambda$ on $\R$, and that $t$ in \eqref{def_X_N} is distributed according to the law $\lambda$. Therefore, $u=\frac{2t}{\pi}$ has density $\sigma=\frac{\pi}{2}\rho(\frac{\pi}{2}\cdot)$.

We now can write $\mathbf{X}_N(t)=\mathbf{X}\!\left(u+ie^{-s},0;\sve{\frac{\gamma}{2}u}{0},\tfrac{\gamma^2}{8}u\right)$ and  proving  Theorem \ref{main-thm} is equivalent to showing that for every bounded, continuous function $h:\mathbb C^2\to\mathbb R$, we have that 

\begin{align}\label{reduced-statement}\lim_{s\to\infty}\int_{\R}h\!\left(\mathbf{X}\!\left(u+ie^{-s},0;\sve{\frac{\gamma}{2}u}{0},\tfrac{\gamma^2}{8}u\right)\right)\sigma(u)\de u=\frac{1}{\mu(\GamG)}\int_{\GamG}h\!\left(\mathbf{X}( \Gamma g)\right)\de\mu(\Gamma g)
\end{align}

Observe that if $f_0,f_1\in\mathcal{S}_\eta$ with $\eta>1$, then $h\circ\mathbf{X}$ is bounded and continuous and, since we are assuming that $\gamma$ is irrational, we can apply Theorem \ref{thmJEQ2} to achieve \eqref{reduced-statement}.

Finally, if $f_0$ and $f_1$ are bounded, Riemann-integrable functions, then we can use a standard approximation argument, analogous to the one used in Lemmata 4.5--4.9 in \cite{PLMS:PLMS0775}. \qed

\begin{remark}\label{remark-independence}
Since the two components of $\textbf{X}$ are both functions of the same random variable ($\Gamma g$ distributed in $\GamG$ according to the normalized Haar measure), it is clear that  the law on $\C^2$ of $\mathbf{X}$ is not the product measure of its two marginals on $\C$. The analogue of Theorem \ref{main-thm} for rational $\gamma$ can also be considered, but the statement of Theorem \ref{thmJEQ2} needs to be modified. In fact, the hororcyle lifts do not equidistribute on the homogeous space $\GamG$, instead they equidistribute on a submanifold of positive codimension of the form $\Gamma_\gamma\backslash G$. In this case the random variable $\Gamma g$ on the right-hand-side of \eqref{statement-limit-thm} would be distributed according the normalized Haar measure on $\Gamma_\gamma\backslash G$. The extension of  Theorem \ref{thmJEQ2}  for rational $\gamma$ is the subject of a joint work in progress with Tariq Osman.
\end{remark}

\section{A class of SUSY partner P\"{o}schl-Teller potentials}\label{section-many-susy-partners}
\begin{figure}[h!]
\begin{center}
\vspace{-4cm}
\includegraphics[scale=.6]{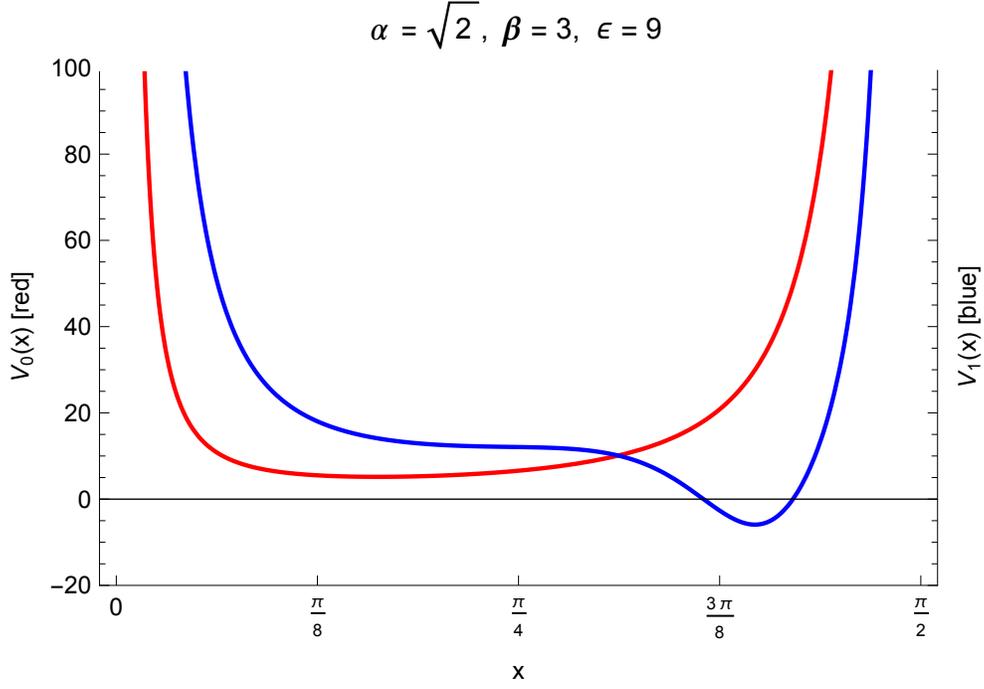}
\vspace{-4cm}
\end{center}
\caption{The potentials $V_0$ and $V_1$ of two isospectral supersymmetric partner Hamiltonians, given by \eqref{def-V0} and \eqref{partner-V_1}, respectively.}
\label{two-potentials}
\end{figure}

In this section we explore two constructions that, given a P\"{o}schl-Teller potential, allows us to construct infinitely many isospectral supersymmetric partners, thus providing us with a plethora of cases in which our Theorem \ref{main-thm} 
can be applied.
Let $H_0$ be the Hamiltonian with potential $V_0$ as in \eqref{def-V0}. Recall that $\gamma=\alpha+\beta$. 
A general solution  to the equation $H_0 u=\epsilon u$ (regardless of boundary conditions) for any $\epsilon>0$ is 
\begin{align}
u_{\epsilon,A,B}(x)=&\sin^\alpha(x)\cos^\beta(x)\left(A\,\, _2F_1\!\left(\frac{\gamma}{2}+\sqrt{\frac{\epsilon}{2}},\frac{\gamma}{2}-\sqrt{\frac{\epsilon}{2}};\alpha+\frac{1}{2};\sin^2(x)\right)\right.\label{eq:u(x)}\\
&\left.+B\,\sin^{1-2\alpha}(x)\,\, _2F_1\!\left(\frac{1+\beta-\alpha}{2}+\sqrt{\frac{\epsilon}{2}},\frac{1+\beta-\alpha}{2}-\sqrt{\frac{\epsilon}{2}};\frac{3}{2}-\alpha;\sin^2(x)\right)\right).\nonumber
\end{align}
Using first-order intertwining operators of the form,
\begin{align}
A=\frac{1}{\sqrt2}\left(\frac{\de}{\de x}+\kappa(x)\right),&&A^\dag=\frac{1}{\sqrt2}\left(-\frac{\de}{\de x}+\kappa(x)\right),\label{eq:AandAdag}
\end{align} Contreras-Astorga and Fern\'andez \cite{MR2515866}, are able to construct a 1-parameter family of partner potentials $V_1$ such that $H_0$ and $H_1$ are isospectral. See also \cite{SUSYQM-Fernandez}.
Specifically
\begin{align}
\label{partner-V_1}
V_1(x)=\frac{\alpha(\alpha+1)}{2\sin^2(x)}+\frac{(\beta-2)(\beta-1)}{2\cos^2(x)}-\left(\log v_\epsilon(x)\right)'',\hspace{.5cm}\alpha>1,\beta>2,
\end{align}
where 
\begin{align}
v_\epsilon(x)&=\frac{u_{\epsilon,1,0}(x)}{\sin^\alpha(x)\cos^{1-\beta}(x)}=\cos^{2\beta-1} (x)_2F_1\!\left(\frac{\mu}{2}+\sqrt{\frac{\epsilon}{2}},\frac{\gamma}{2}-\sqrt{\frac{\epsilon}{2}};\alpha+\frac{1}{2};\sin^2(x)\right),
\end{align}
and $\epsilon<E_0=\frac{\mu^2}{2}$ is an arbitrary real parameter. The corresponding eigenfunctions for $H_1$ are
\begin{align}
\psi_{1,n}=\frac{A^\dag\psi_{0,n}}{\sqrt{E_n-\epsilon}}
\end{align}
where  $\kappa(x)=(\log u_{\epsilon,1,0}(x))'$  in \eqref{eq:AandAdag}.
An example of such a potential $V_1$ for $\alpha=\sqrt{2}$ and $\beta=3$ is shown in Figure \ref{two-potentials}, along with the supersymmetric partner $V_0$.

 In the same paper \cite{MR2515866} the authors also use second-order intertwining operators $B,B^\dag$ of the form so that $H_1B^\dag=B^\dag H_0$, where
\begin{align}
B^\dag=\frac{1}{2}\left(\frac{\de^2}{\de x^2}-\eta(x)\frac{\de}{\de x}+\theta(x)\right)\label{eq:Bdag}
\end{align}
to construct a 2-parameters family of partner potentials $V_1$ such that $H_0$ and $H_1$ are isospectral. Specifically
\begin{align}
V_1(x)=\frac{(\alpha+1)(\alpha+2)}{2\sin^2(x)}+\frac{(\beta-3)(\beta-2)}{2\cos^2(x)}-\left(\log \mathcal W(x)\right)'',\hspace{.5cm}\lambda>1,\nu>3,
\end{align}
where 
\begin{align}
\mathcal W(x)&=\frac{W(u_{\epsilon_1,1,0},u_{\epsilon_2,1,0})}{\sin^{2\alpha+1}(x)\cos^{3-2\beta}(x)}, 
\end{align}
$W(f,g)=f'g-fg'$ denotes the Wronskian of $f$ and $g$,  
$u_{\varepsilon_i,1,0}$ is as in \eqref{eq:u(x)} for $i=1,2$ and $\epsilon_1,\epsilon_2$ are real parameters such that
$E_l<\epsilon_2<\epsilon_1<E_{l+1}$ for some $l\geq 0$.
In this case the normalized eigenfunctions of $H_1$ are
\begin{align}
\psi_{1,n}=\frac{ B^\dag \psi_{0,n}}{\sqrt{(E_n-\epsilon_1)(E_n-\epsilon_2)}},
\end{align}
where in \eqref{eq:Bdag} we have
$\eta=(\log(W(u_{\epsilon_1,1,0},u_{\epsilon_2,1,0})))'$, $\theta=\frac{\eta'}{2}+\frac{\eta^2}{2}-2V_0+d$, and $d=\epsilon_1+\epsilon_2$. 

In all these classes of supersymmetric partner potentials, our Theorem \ref{main-thm} applies, provided $\gamma$ is irrational


%
\section*{Acknowledgments}
The author would like to thank Tariq Osman for many valuable discussions on the equidistribution of horocycle lifts, and the anonymous referee for their helpful remarks. Furthermore, the author acknowledges the support 
of the NSERC Discovery Grant \emph{``Statistical and Number-Theoretical Aspects of Dynamical Systems''}.

\bibliographystyle{plain}
\bibliography{SUSYPT-bibliography}

\end{document}